\documentclass[12pt]{iopart}
\usepackage{graphics}
\begin{document}

\title{Dynamics and freeze-out of hadron resonances at RHIC}

\author{Marcus Bleicher, Horst St\"ocker}

\address{ Institut f\"ur Theoretische Physik,
J.~W.~Goethe Universit\"at, 60054 Frankfurt am Main,
Germany}

\begin{abstract}
Yields, rapidity and transverse momentum spectra of 
$\Delta^{++}(1232)$, $\Lambda(1520)$, $\Sigma^\pm(1385)$ and  
the meson resonances $K^0(892)$, $\Phi$, $\rho^0$ and $f_0(980)$
are predicted. Hadronic rescattering leads to a suppression of
reconstructable resonances, especially at low $p_\perp$.
A mass shift of the $\rho$ of 10~MeV is obtained from the 
microscopic simulation, due to late stage $\rho$ formation in the
cooling pion gas.
\end{abstract}

\maketitle

Strange particle yields and spectra are key  probes
to study excited nuclear matter and to detect
the transition of (confined) hadronic matter to
quark-gluon-matter (QGP)\cite{qgpreviews,rafelski}.
The relative enhancement of strange and  multi-strange
hadrons, as well as hadron ratios in central
heavy ion collisions with respect to peripheral or proton
induced interactions have been suggested as a signature
for the transient existence of a QGP-phase \cite{rafelski}.
A main difficulty in the interpretation of the available data is that
the observed final state
hadrons carry relatively little information about their primordial sources.
Most of the hadrons had been subject to many secondary
interactions and are strongly influenced
by the decays of high mass resonances.

Recently, first direct experimental information on unstable 
particle emission in nucleus-nucleus reactions has been reported. 
At SPS \cite{NA49Res}, 
the $\Phi$, $\overline{K^0}(892)$ and $\Lambda(1520)$ have been observed, 
at the full RHIC energy first data on the $\Phi$, ${K^0}(892)$, 
$f_0$ and $\rho$ mesons and the $\Lambda(1520)$ and 
$\Sigma(1385)$ baryon resonances are available \cite{STARkstar}.
These data are obtained by the 
reconstruction of final state hadrons. Thus, in contrast to 
the study of $\rho$ mesons by di-leptons (which are penetrating probes), 
it yields information mostly from the later stages of the reaction.
Therefore, the question of the existence of
such resonance states in the hot and dense environment is still
not unambiguously answered. E.g. hyperon resonances are
expected to dissolve at high energy densities \cite{Lutz:2001dq}
and also the $\rho$ is expected to ''melt'' in the medium \cite{rapp}. 
However, especially for the strange baryon resonances it is expected
that late stage regeneration of the resonance is suppressed,
therefore there yield might still carry information whether they
do exist in the hot and dense region. 
It is therefore of utmost importance to 
study the cross section of hyperon resonance production at
different beam energies and centralities to see whether a
specific suppression threshold is present. 

To study a possible suppression of resonances
and its origin,  the Ultra-relativistic Quantum Molecular 
Dynamics model (UrQMD 1.2) is used \cite{urqmdmodel}.
This microscopic transport approach is based on the
covariant propagation of constituent quarks and di-quarks accompanied
by mesonic and baryonic degrees of freedom. 
The leading hadrons of the fragmenting strings contain the valence-quarks
of the original excited hadron and represent a simplified picture
of the leading (di)quarks of the fragmenting string.
The elementary hadronic interactions are modelled according to
measured cross sections and angular distributions. If the cross
sections are not experimentally known, detailed balance is employed
in the energy range of resonances. The partial and total decay widths
are taken from the Particle Data Group. Presently, in-medium modifications
of the particle properties, i.e. temperature and density dependent masses
and decay widths, are not included in the model. Thus, it serves as a
benchmark calculation and allows in comparison to experimental data
to identify a  possible anomalous behaviour in the resonance dynamics.
In these proceedings, we will focus on nucleus-nucleus collisions at 
RHIC energies. For a discussion of resonance production at SPS 
and resonance excitation functions within the same approach
the reader is referred to Refs.  \cite{Bleicher:2002dm,Bleicher:2002rx}.
An exploration of the freeze-out conditions in heavy ion reactions 
from the perspective of a refined thermal model can be 
found in \cite{Rafelski:2001hp,Torrieri:2001tg,Torrieri:2002jp}.
\begin{figure}
\resizebox*{!}{0.35\textheight}{\includegraphics{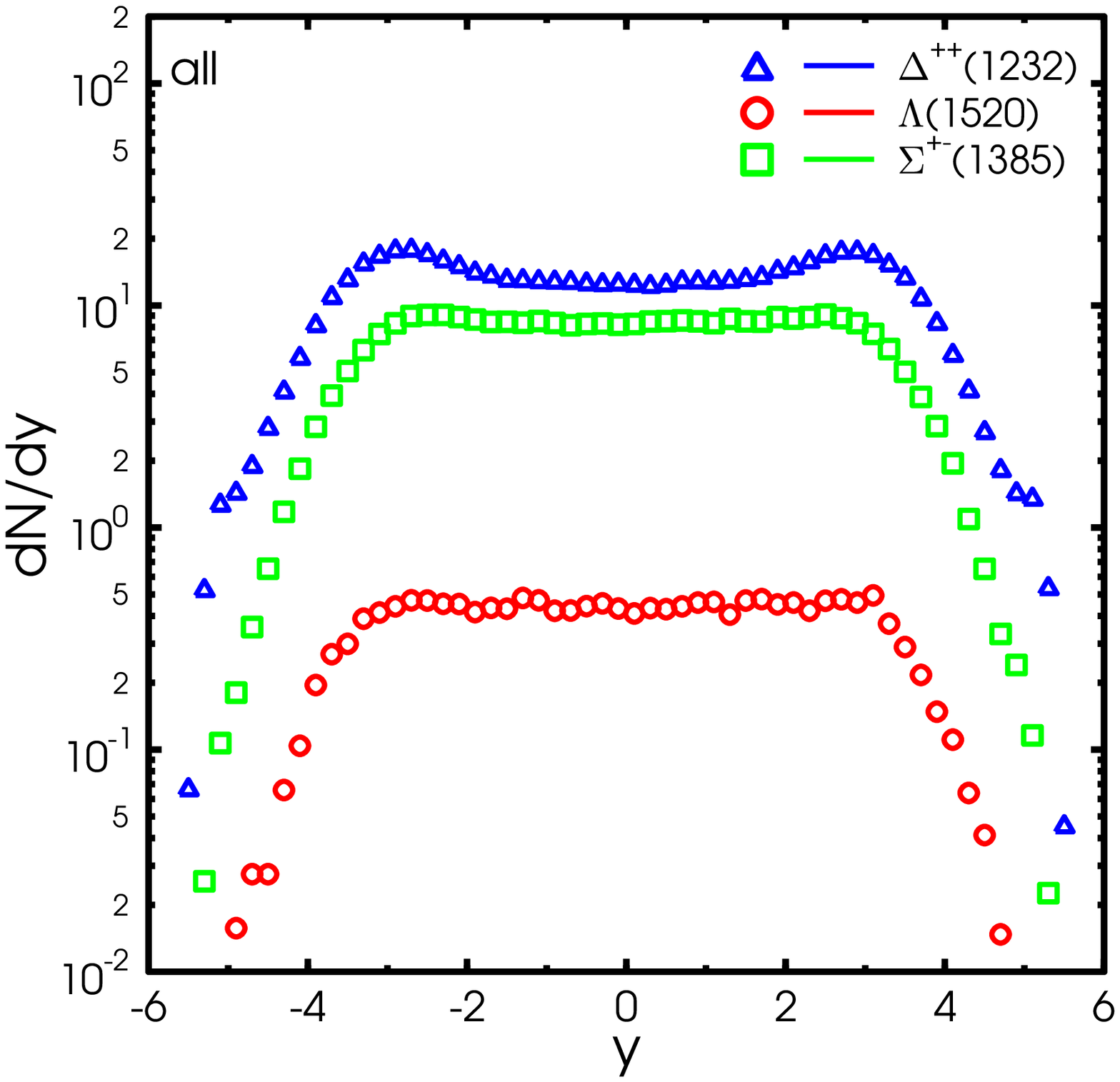}
\includegraphics{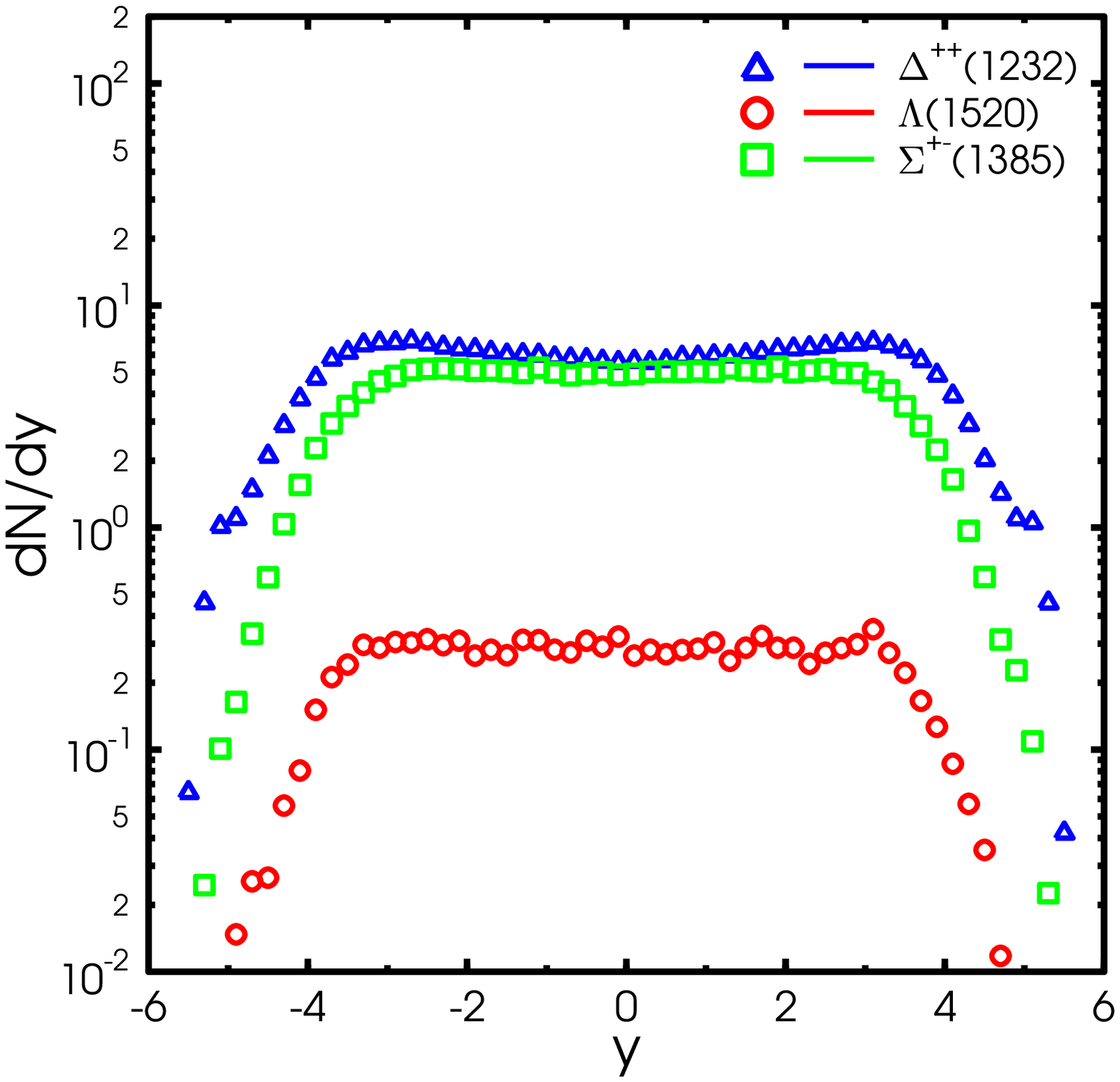}}
\caption{Rapidity densities of baryon resonances in central Au+Au
reactions at $\sqrt s = 200$~AGeV. Left: All baryons resonances
that decay are shown. Right: Only those baryon resonances are shown that can
 (in principle) be reconstructed by experiment because their decay products
did not interact after the decay of the resonance.
\label{dndybaryon}}
\end{figure}
\begin{figure}
\resizebox*{!}{0.35\textheight}{\includegraphics{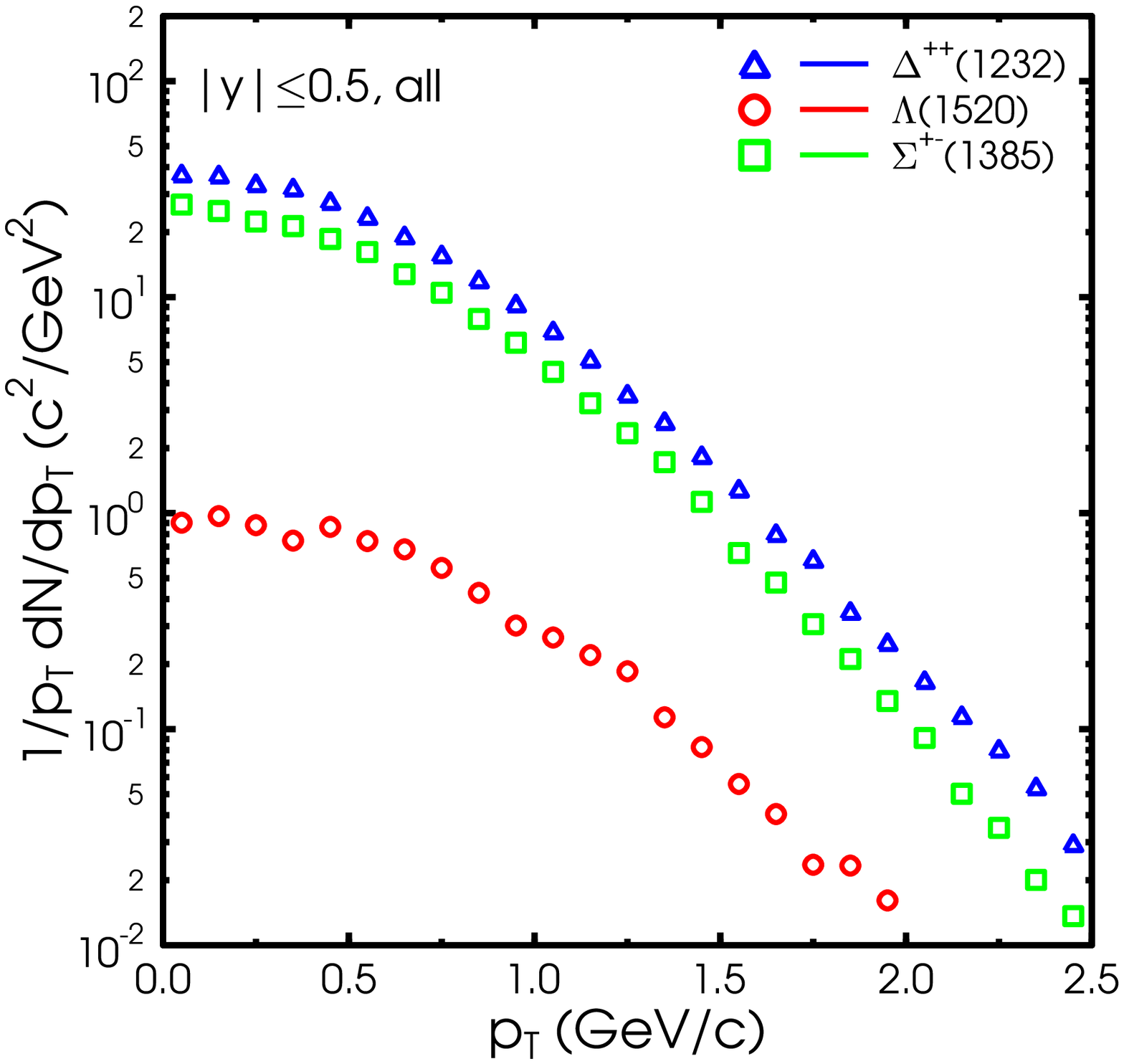}
\includegraphics{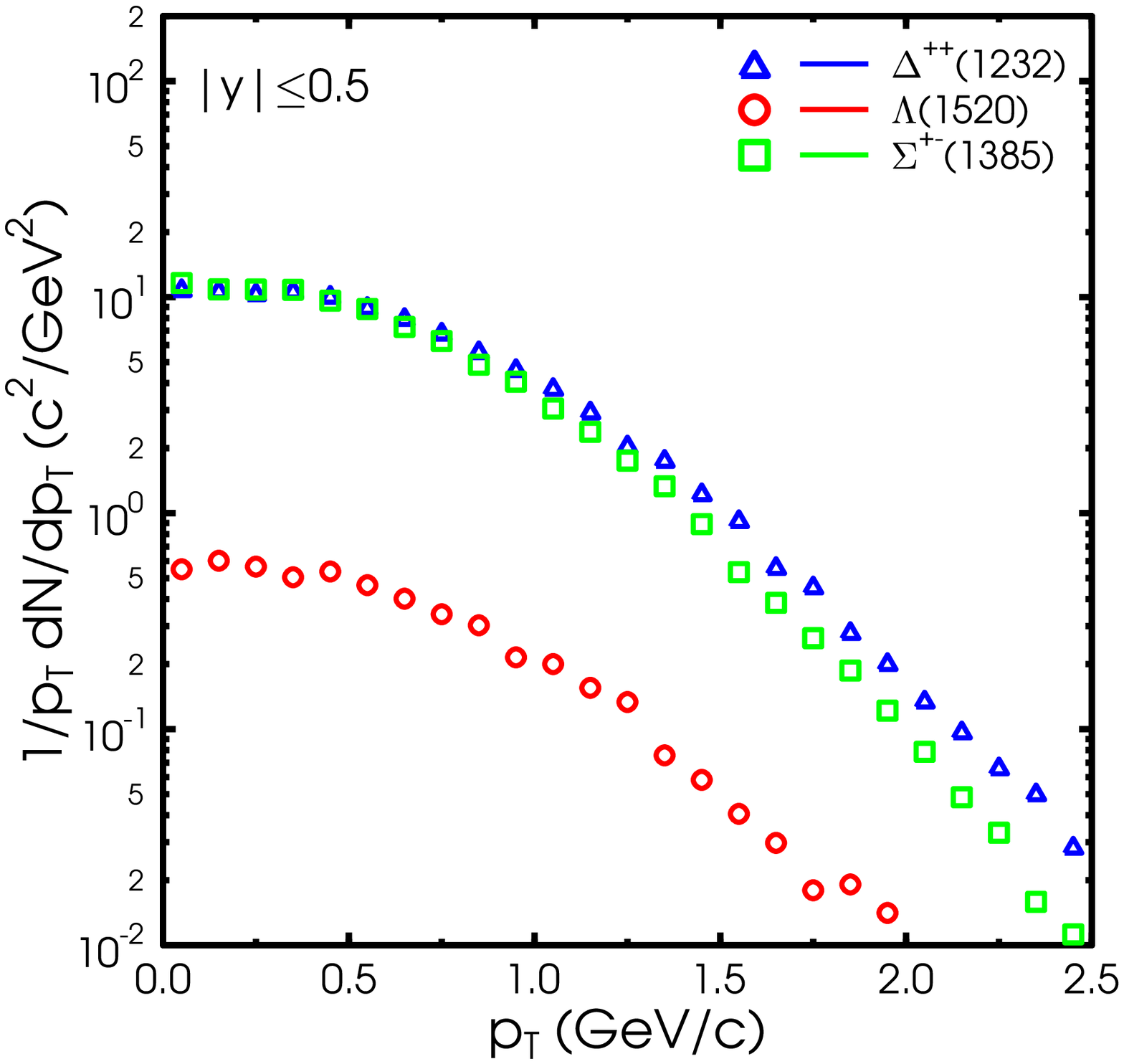}}
\caption{Transverse momentum spectra of baryon resonances in central Au+Au
reactions at $\sqrt s = 200$~AGeV. Left: All baryons resonances
that decay are shown. Right: Only those baryon resonances are shown that can
 (in principle) be reconstructed by experiment.
\label{dndptbaryon}}
\end{figure}
\begin{figure}
\resizebox*{!}{0.35\textheight}{\includegraphics{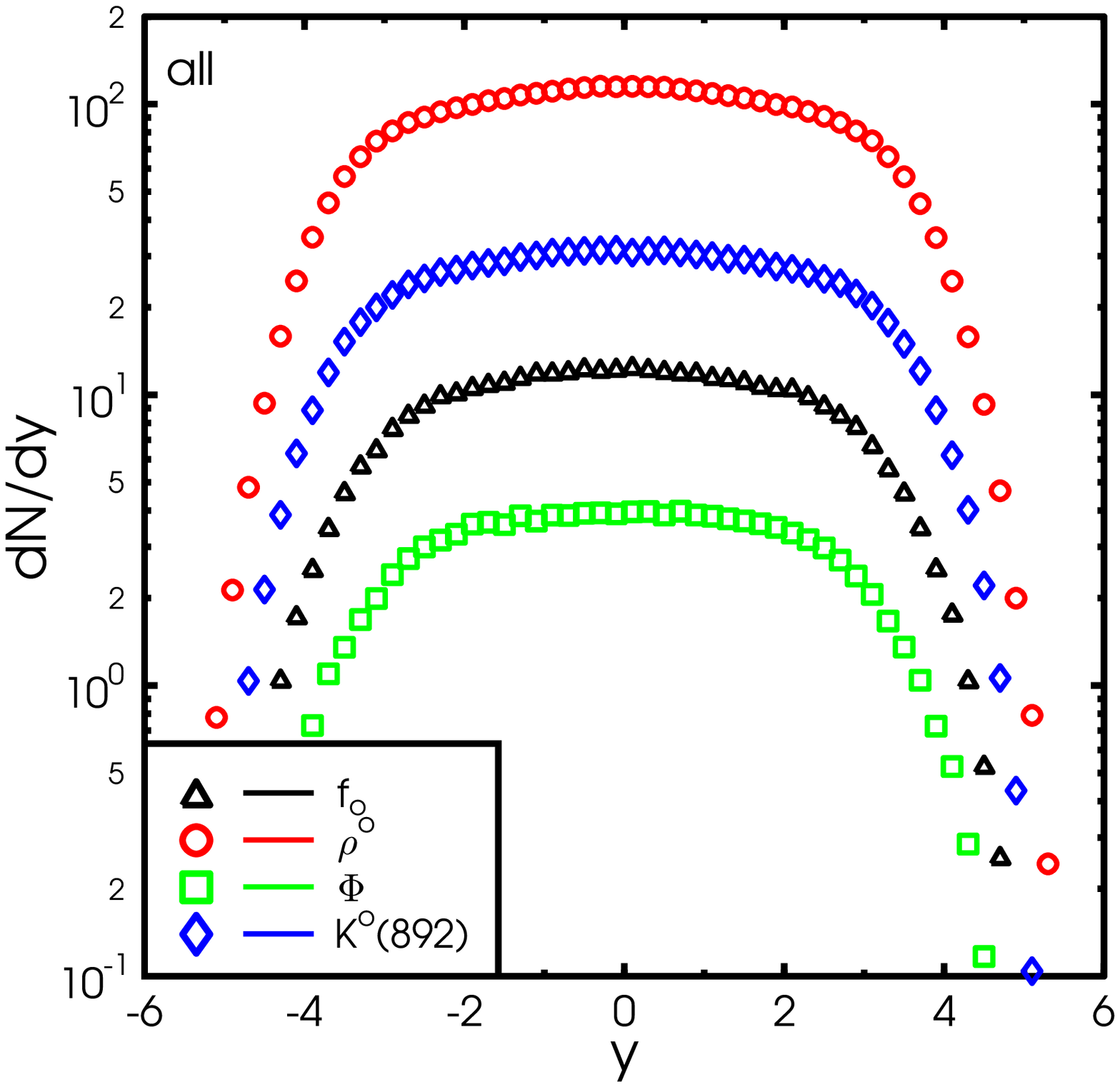}
\includegraphics{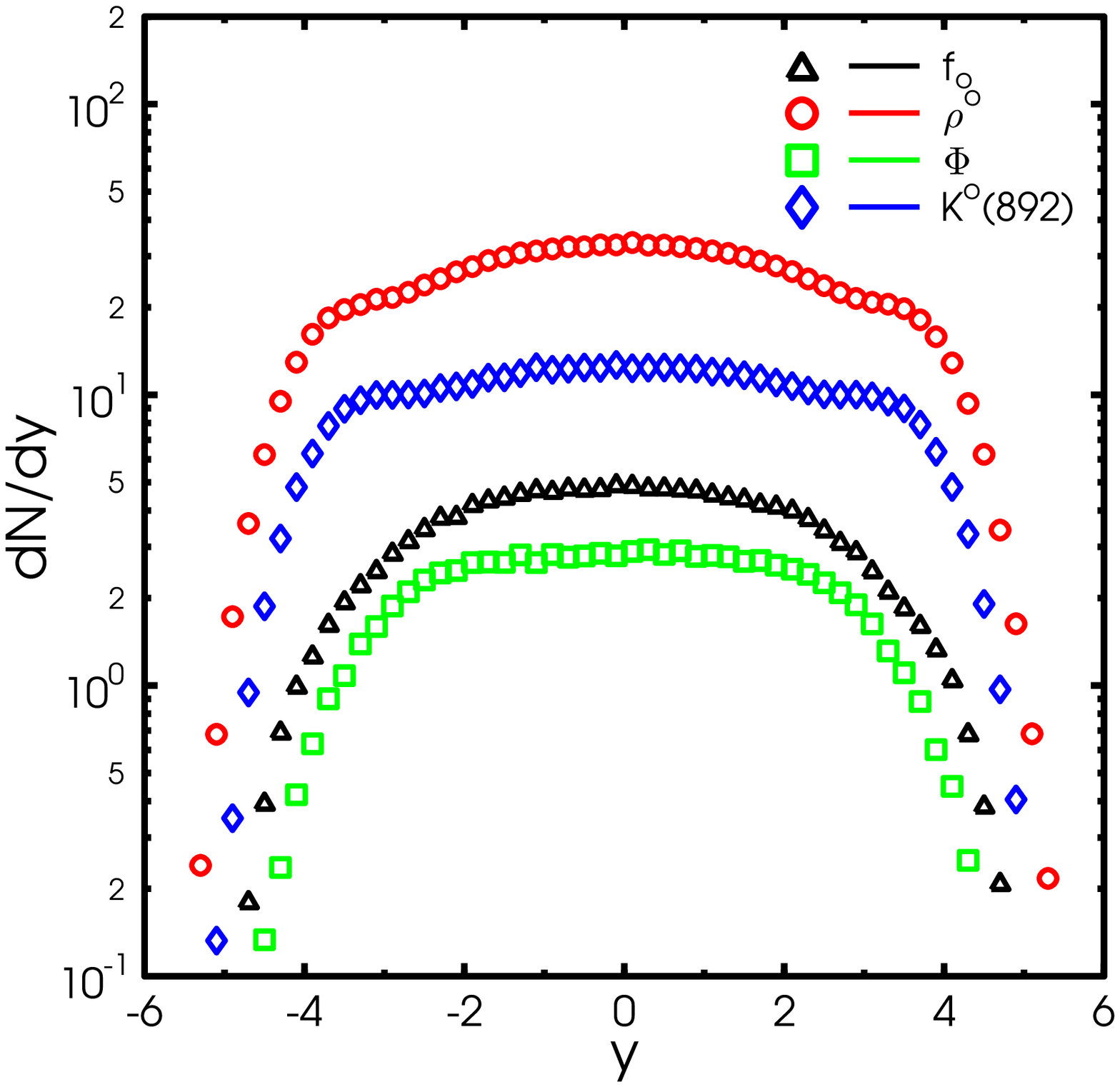}}
\caption{Rapidity densities of meson resonances in central Au+Au
reactions at $\sqrt s = 200$~AGeV. Left: All meson resonances
that decay are shown. Right: Only those meson resonances are shown that can
 (in principle) be reconstructed by experiment because their decay products
did not interact after the decay of the resonance.
\label{dndymeson}}
\end{figure}
\begin{figure}
\resizebox*{!}{0.35\textheight}{\includegraphics{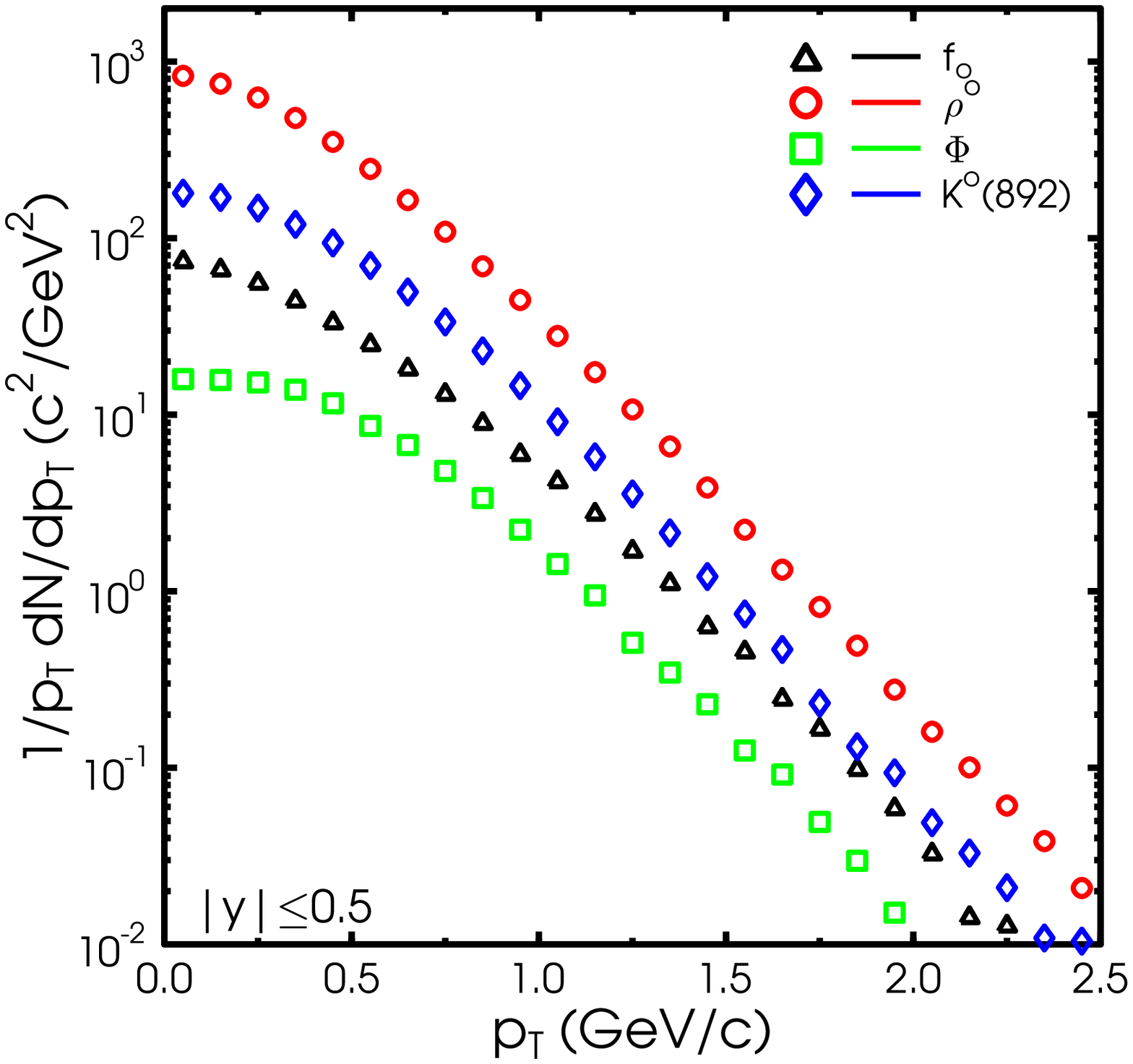}
\includegraphics{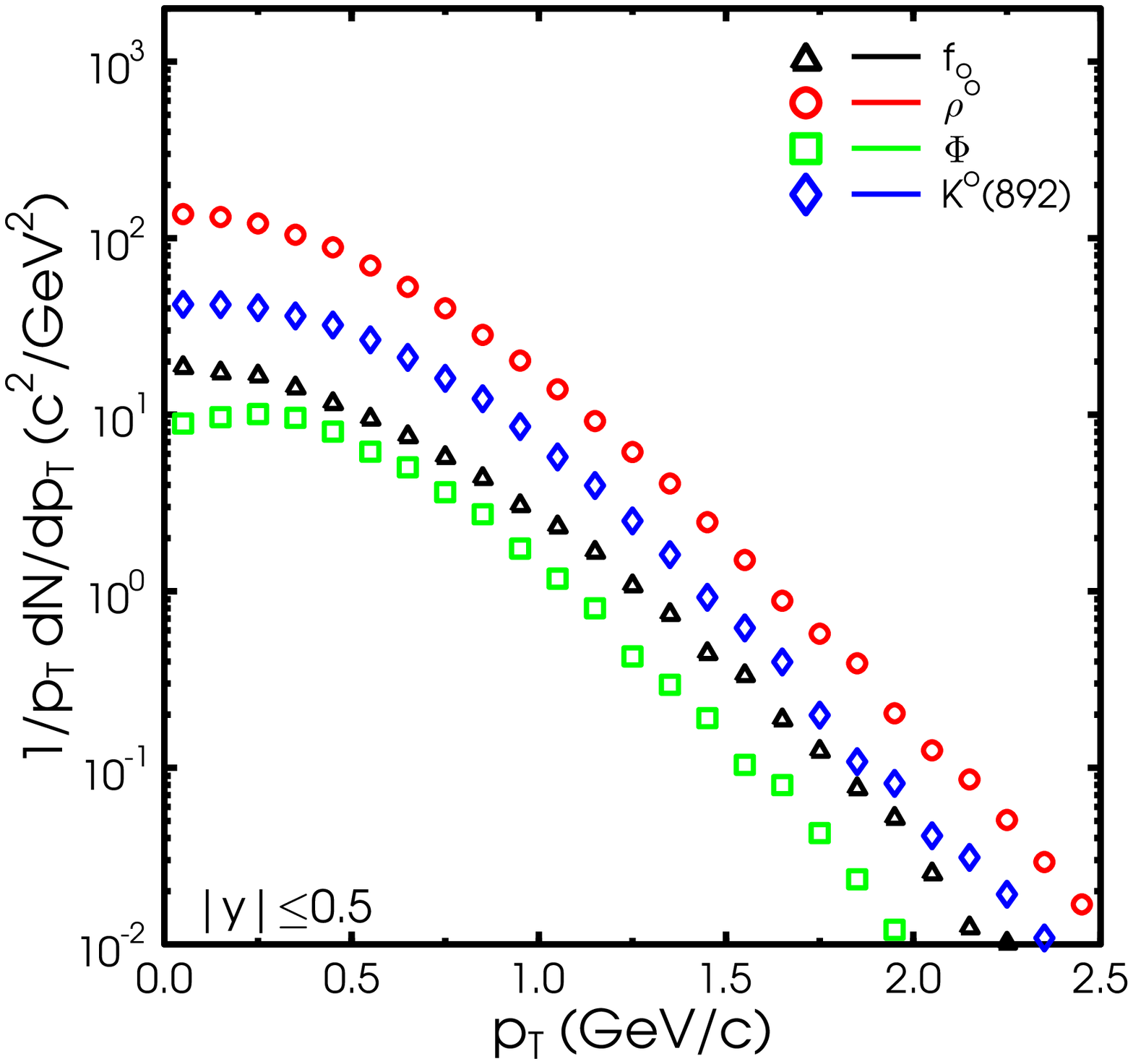}}
\caption{Transverse momentum spectra of meson resonances in central Au+Au
reactions at $\sqrt s = 200$~AGeV. Left: All meson resonances
that decay are shown. Right: Only those meson resonances are shown that can
 (in principle) be reconstructed by experiment.
\label{dndptmeson}}
\end{figure}

In general, a nucleus-nucleus collision evolves through
three distinct stages: (I) The initial stage is characterised by
high scattering rates and deposits a large amount of (non-thermalized) 
energy around central rapidities.
(II) Either partonic or hadronic ''cooking'' may equilibrate the 
system until the individual inelastic scattering rates drop as  
the system cools down - usually called the chemical freeze-out.
(III) Elastic and pseudo-elastic  interactions (e.g. $\pi\pi \rightarrow \rho 
\rightarrow\pi\pi$) dominate the final stage of the
collision, ending with the final break-up of the ''fireball'' (kinetic 
freeze-out).
The spectra and abundances of meson and baryon resonances 
are well suited to study the break-up dynamics of the source between
chemical and kinetic freeze-out.

Let us start by an overview of the rapidity and transverse momentum
spectra in central Au+Au collisions at $\sqrt s = 200$~AGeV.
Fig. \ref{dndybaryon} shows the rapidity densities of the
$\Delta^{++}(1232)$, $\Lambda(1520)$ and the charged $\Sigma^\pm(1385)$
states in central Au+Au reactions at $\sqrt s = 200$~AGeV. 
The left part of Fig. \ref{dndybaryon}  
includes all decaying baryon resonances in the spectrum, 
while the right part of Fig. \ref{dndybaryon} shows the finally
observable spectrum of resonances. Different to the
experiment, were a mixed event technique is used to obtain an
invariant mass spectrum at any given $p_\perp$ and rapidity, 
in the present study we define a resonance as observable 
if its decay products do not interact further after the decay of 
the resonance. This procedure has the advantage that for each 
individual observed 
resonance information on its origin and decay time can be obtained from
the model.
However, this definition might yield a slightly stronger suppression 
of resonances than the mixed event technique, because it assumes that any 
momentum transfer  to the decay products will result in
a complete loss of the resonance signal in the invariant mass spectrum of
the two correlated hadrons.

Fig. \ref{dndptbaryon} shows the transverse momentum spectra 
of $\Delta^{++}(1232)$, $\Lambda(1520)$ and the charged $\Sigma^\pm(1385)$
states in central Au+Au reactions at $\sqrt s = 200$~AGeV. 
Here, one observes a significant reduction of the low $p_\perp$ part of
the spectrum, while the high $p_\perp$ part remains essentially unchanged.
This apparent heating of the resonance spectrum is partly due to
flow, but mostly due to the higher absorption probability for the
decay products of low $p_\perp$ resonances because they decay already 
inside the hot and dense region.

Figures \ref{dndymeson} and \ref{dndptmeson} show the
same information for the meson resonances $K^0(892)$, $\Phi$, $\rho^0$ and the
neutral $f_0(980)$. The left parts of the figures include all 
resonances as they decay, while the right parts show only those resonances
that are reconstructable from an invariant mass spectrum of 
final state hadrons. 
I.e. the left parts of Figs. \ref{dndymeson} and \ref{dndptmeson}
can be studied in the di-lepton channel (e.g. in 
the $\rho^0$ decay, normalised by the branching ratio $\Gamma_{\rho\rightarrow l^+l^-}$), while the right parts show predictions for the 
spectra by reconstruction of  $K\pi$, $\overline KK$ and $\pi\pi$ hadron
correlations.

Especially the $\rho$ meson with its two distinct decay modes ($e^+e^-$ vs.
$\pi\pi$), which are both accessible by experiment, allows to explore 
the dynamics and thermodynamic properties of the system. 
Therefore, we will now focus on the freeze-out and decay 
of $\rho$ mesons.

\begin{figure}
\resizebox*{!}{0.35\textheight}{\includegraphics{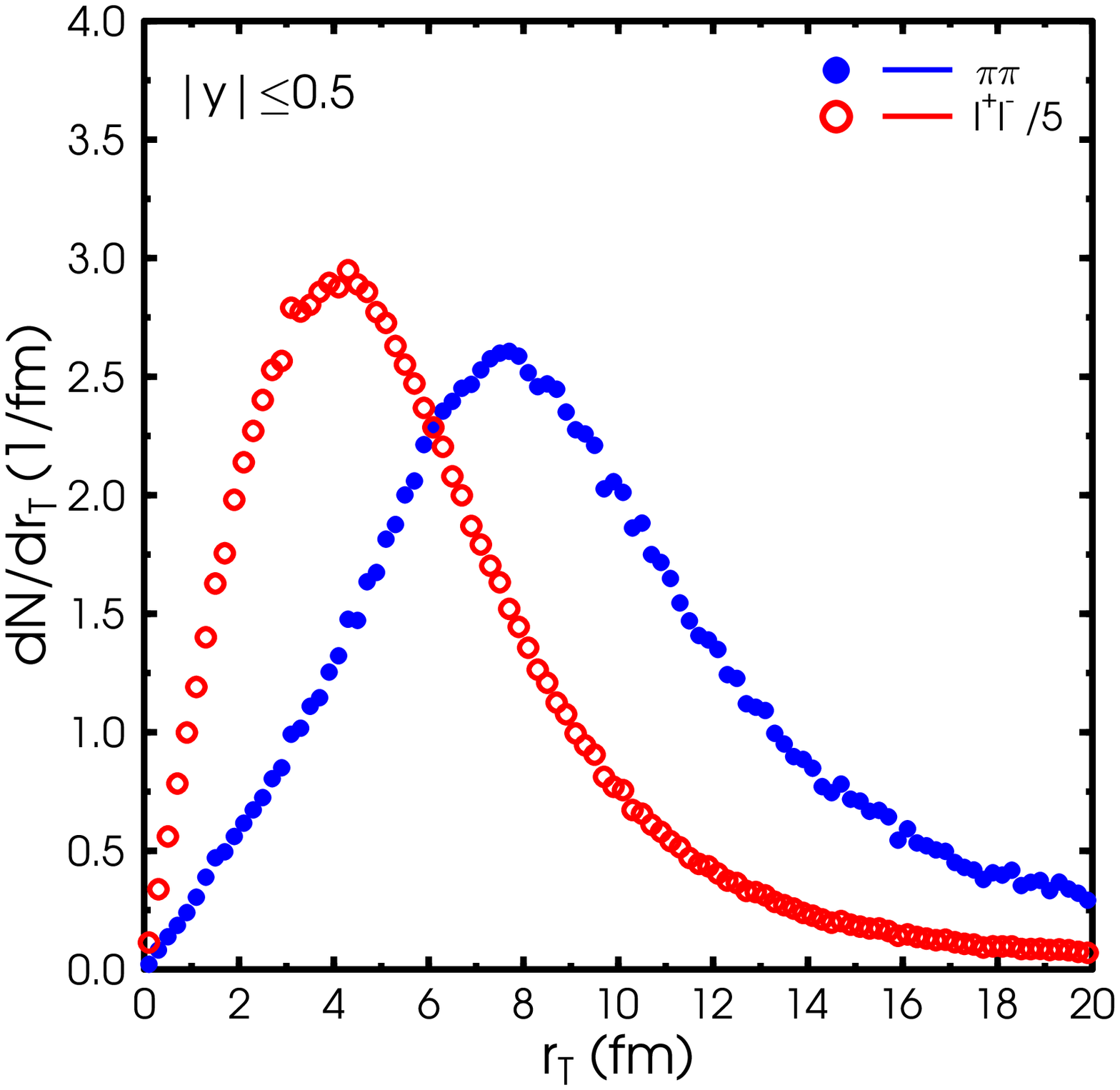}
\includegraphics{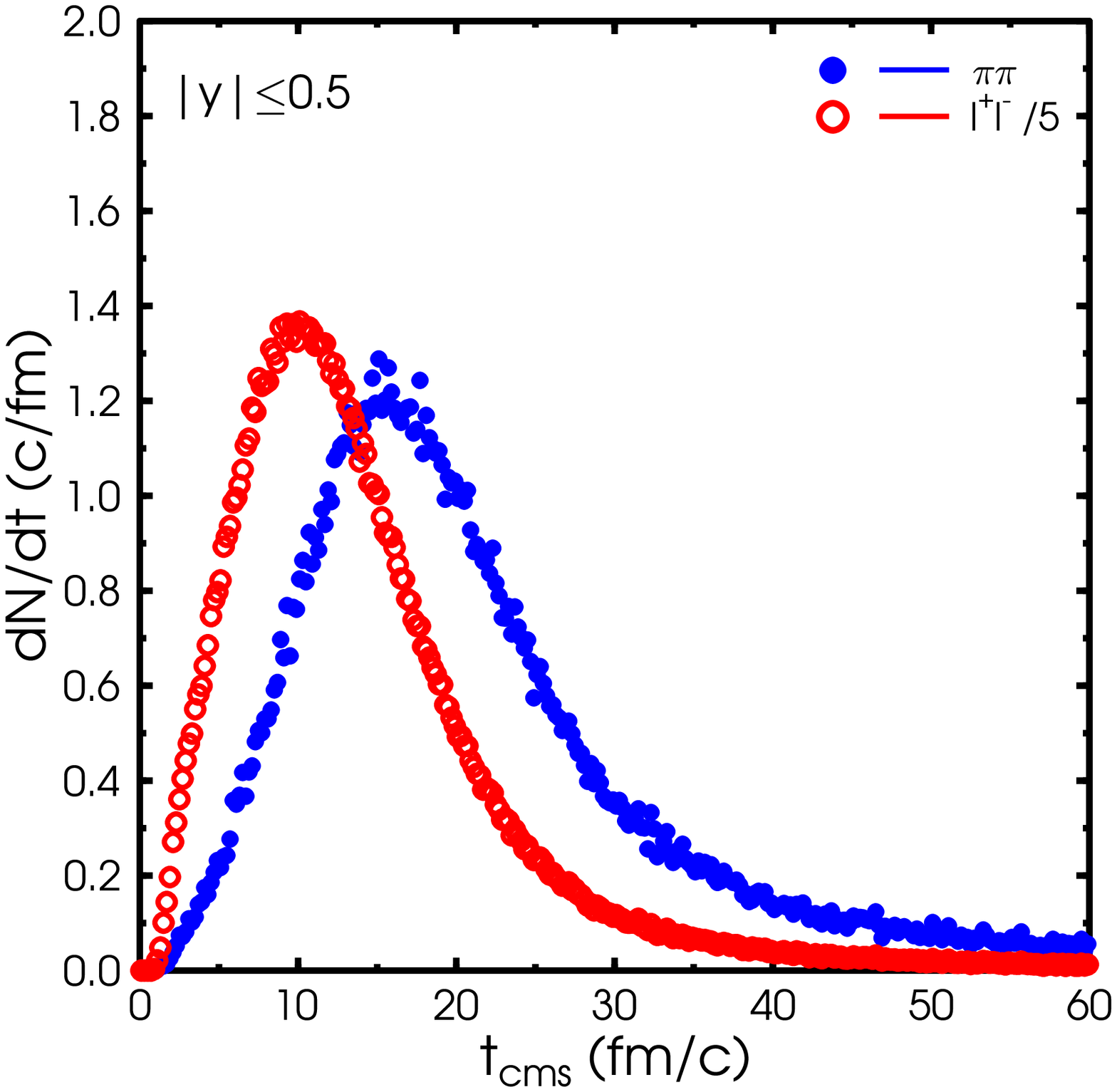}} 
\caption{Time and transverse position of the $\rho$ decay around 
midrapidity ($|y|\le 0.5$)
in central Au+Au reactions at $\sqrt s = 200$~AGeV.
Closed symbols denote decaying $\rho$ mesons reconstructable
in the pion channel, open symbols show the di-lepton channel (normalised
by the branching ratio). 
Left: Transverse radii at which the $\rho^0$ mesons decay 
Right: Center of mass times at which the $\rho^0$ mesons decay. 
\label{dndrtrho}}
\end{figure}

In Figure \ref{dndrtrho} (left) the transverse radii at 
which  $\rho^0$ mesons decay in central Au+Au reactions at 
the full RHIC energy are shown.  Open symbols depict the
di-lepton channel (divided by 5 for better visibility and 
normalised by the respective branching ratio), 
full symbols show the $\rho^0$'s reconstructable in the
pion channel. One clearly observes that the $\pi\pi$ channel 
is mainly sensitive to the outer layers of the interaction region. Here,
the emission peaks at $r_\perp = 8$~fm. Studying the $\rho$ in the
di-lepton channel however, reveals information mainly from the interior
of the reaction zone, with a peak around $r_\perp = 4$~fm. 

The same behaviour is reflected in the decay times of the $\rho$'s 
at midrapidity (see Fig. \ref{dndrtrho} (right)). While the
pion channel automatically triggers on the final stages of the
interaction (full symbols) with peak emission at $t = 15$~fm/c,
the di-lepton channel, shown as open symbols, probes 
the density and temperature evolution around 10~fm/c after 
the begin of the interaction. 

It is interesting to note that the $\rho$ yield as observed in the 
pion channel is dominated by regenerated $\rho$'s. I.e. since the life time
of the $\rho$ state is $\sim 1.5$~fm/c, the finally observed $\rho$ mass
spectrum and yield is sensitive to the temperature and densities of the
pions near the final break-up of the system.

Indeed, preliminary STAR data \cite{fachini} seems to indicate a 
mass shift in the $\rho$ if reconstructed by pion correlations.
A first rough estimate of the $\rho$ mass shift due to a cooling down
of the pion gas can be obtained
by folding the Breit-Wigner distribution of the $\rho$ mass with
a thermal distribution of temperature $T$:
\begin{equation}
\frac{{\rm d}N_\rho}{{\rm d}^3p \,{\rm d}m} 
\sim {\rm exp}[-E_\rho/T] A(m)\quad;\quad
A(m) \sim \frac{\Gamma_\rho}{(m-m^{\rm peak}_\rho)^2 + (\Gamma_\rho/2)^2}\quad,
\end{equation}
for simplicity, phase space factors that modify the tails of the
mass distribution are not included here. Thus, the finite
width of the $\rho$ results in a temperature
dependence of the peak mass of the $\rho$ as given by
\begin{equation}
\frac{{\rm d}N_\rho}{{\rm d}m} \sim m^2 T K_2(m/T) A(m)\quad.
\end{equation}
For decoupling temperatures of interest, 100~MeV$\le T \le$ 150~MeV,
this leads to a  downward mass shift of the $\rho$ peak by about 10-20~MeV.
Note that in contrast to this simple estimate, the model calculations 
always includes  phase space corrections
and mass dependent decay widths for resonances.

\begin{figure}
\resizebox*{!}{0.35\textheight}{
\includegraphics{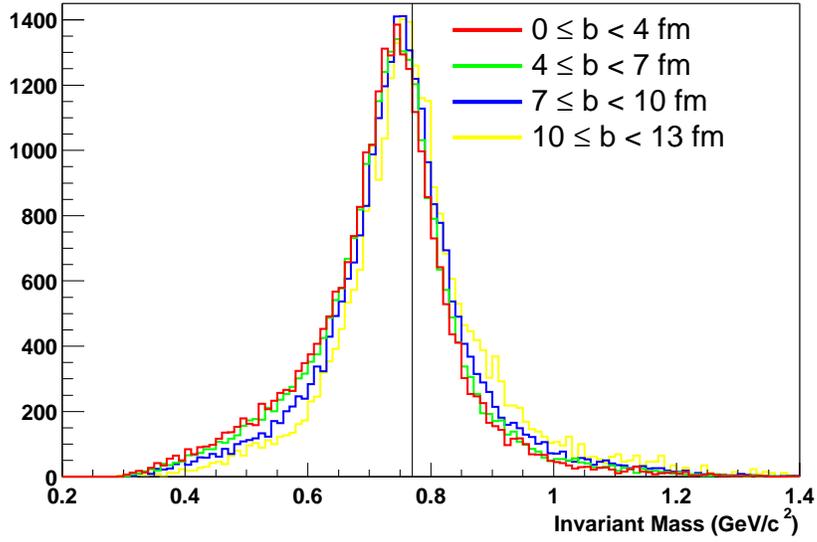}} 
\caption{Centrality dependence of the mass distribution 
of $\rho^0$ mesons being reconstructed in
the $\pi\pi$ channel in minimum biased Au+Au reactions 
at $\sqrt s = 200$~AGeV. 
 \label{UrqmdRhoMassImpPar}}
\end{figure}

Let us compare these estimates for the $\rho$ mass shift to the results 
obtained in the present model.
In Figure \ref{UrqmdRhoMassImpPar} the centrality dependence of
the $\rho$ mass spectrum is shown for minimum biased Au+Au reaction 
at RHIC. When going from peripheral to central collisions one finds a 
small mass shift in the $\rho$ distribution on the order of 10~MeV.
This indicates, that the increased rescattering in central collisions
and the regeneration of the $\rho$ due to pion rescattering in 
the late stage of the reaction can indeed lead to a shift in the 
$\rho$ mass.

\begin{figure}
\resizebox*{!}{0.35\textheight}{
\includegraphics{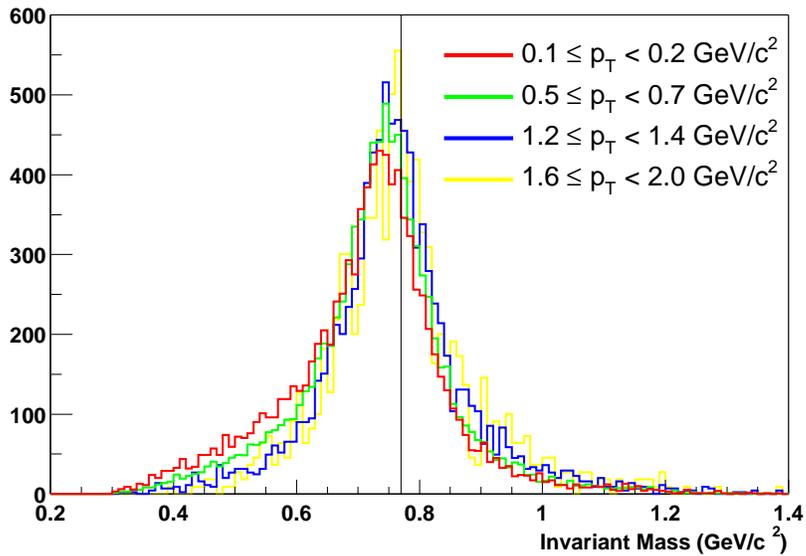}} 
\caption{Transverse momentum dependence of the mass distribution 
of $\rho^0$ mesons  being reconstructed in
the $\pi\pi$ channel in central Au+Au reactions at $\sqrt s = 200$~AGeV. 
 \label{UrqmdRhoMassPt}}
\end{figure}

To explore this behaviour further, Fig. \ref{UrqmdRhoMassPt} shows the 
mass distribution of the $\rho^0$ mesons being reconstructed in
the $\pi\pi$ channel in central Au+Au collision at $\sqrt s = 200$~AGeV
for different $p_\perp$ windows.
Here, the present model predicts a strong transverse momentum dependence
of the $\rho$ mass shift. At low $p_\perp$ the mass shift 
exceeds 30~MeV, while it vanishes above 1.5~GeV in transverse momentum.
This observation also supports, the interpretation of the $\rho$ mass shift
in the model as a signal for late stage formation of $\rho$'s due to $\pi\pi$
scattering. 

Comparing to the preliminary STAR data \cite{fachini} which yields 
downward shift of the $\rho$ mass  on the order of 60-70~MeV, 
the observed mass
shift in the simulation is too small. It has been suggested that the
additional mass shift can be interpreted as the onset of 
Brown-Rho scaling and various other in-medium modifications of the $\rho$ 
not included in the present calculation \cite{Shuryak:2002kd,Kolb:2003bk}.

In conclusion, a microscopic transport approach is used to study
hadron resonance production in Au+Au reactions at $\sqrt s=200$~AGeV.
Rapidity and transverse momentum spectra of reconstructable (strange)
meson  and baryon resonances are predicted.
The effects of hadronic rescattering on the resonance yields and
spectra are discussed.
Special emphasise is put on the $\rho$ mesons, here a comparison
between the di-lepton channel and the $\pi\pi$ channel might allow
to measure the length of the kinetic rescattering stage in nucleus-nucleus
collisions: the di-lepton channel is sensitive to the
intermediate stage of the reaction, while the $\pi\pi$ channel yields
information about the kinetic freeze-out stage close to the break-up 
of the hadronic system.
The hadron dynamics near the break-up leads to a mass reduction 
of the $\rho$ by
10-20~MeV in central reactions most pronounced at low $p_\perp$. 
The mass reduction vanishes towards peripheral 
reactions and high $p_\perp$.

\noindent
\section*{Acknowledgements}
M.B. wants to thank the STAR group at BNL and the Yale University 
for kind hospitality while working on parts of this manuscript.
Fruitful discussion with Patricia Fachini, Zhangbu Xu, Christina Markert,
J\"org Aichelin and Edward Shuryak are gratefully acknowledged.
This work used computational resources provided by NERSC and the
Center for Scientific Computing at Frankfurt (CSC).

\end{document}